\documentclass{ws-ijmpa}
\makeatletter
\newcommand{\preprint}[1]{%
  \let\ws@old@clinebuf\@clinebuf
  \def\@clinebuf{\noindent\hfill\textsf{#1}\par\vspace{4pt}\ws@old@clinebuf}%
}
\makeatother
\preprint{\textsf{FERMILAB--PUB--26--0317--T}}

\usepackage[super]{cite}
\usepackage{xcolor}
\usepackage[verbose,hypertexnames=false]{hyperref}
\hypersetup{colorlinks=false,allbordercolors=blue,pdfborderstyle={/S/U/W 1}}

\definecolor{IITred}{rgb}{0.5,0.05,0.05}
\definecolor{Dgreen}{RGB}{0,96,0}
\definecolor{Dcyan}{cmyk}{.96,0,0,.4}
\definecolor{IITblue}{rgb}{0.05,0.05,0.8}

    \newcommand{\m}{\hbox{ m}}
    \newcommand{\mss}{\hbox{ m/s$^2$}}
  \newcommand{\fss}{\hbox{ ft/s$^2$}}

\newcommand{\jpsi}{\ensuremath{J\!/\!\psi}}

\DeclareMathOperator{\sech}{\mathrm{sech}}

\begin{document}

\markboth{Chris Quigg}{Great Solitary Wave to Force between Quarks}

%
\catchline{}{}{}{}{}
%

\title{From the Great Wave of Translation to the Force between Quarks}

\author{Chris Quigg}

\address{Fermi National Accelerator Laboratory, P.O. Box 500, Batavia, Illinois 60510 USA
\\ quigg@fnal.gov \quad {ORCID iD \href{https://orcid.org/0000-0002-2728-2445}{0000-0002-2728-2445}}
}

\maketitle

%
\begin{abstract}
\emph{\hspace*{\fill}To Gerard 't Hooft on the occasion of his 80th birthday, with admiration.\hspace*{\fill}}

\vspace*{6pt}
The chance observation of a novel traveling wave in a canal led over time to the formulation of a nonlinear wave equation---the Korteweg--de Vries equation---that describes strikingly robust disturbances now called solitons. The figure of an isolated soliton corresponds to a reflectionless potential that supports a single bound state in the one-dimensional Schr\"{o}dinger equation. An appropriate combination of individual solitons yields a symmetric reflectionless potential that supports multiple bound states. Thus, the KdV equation opens the path to solving the inverse scattering problem for a collection of bound states. Applied to the quarkonium spectra, this formalism allows the construction of reflectionless approximations to the  confining potentials that account for the force between quarks, and to tests of the flavor-independence of the interquark interaction.

\end{abstract}



\section{Scott Russell's Observation \label{sec:transwave}}	
Many notable scientific developments spring from observation that blossoms into exploration,  matures into lively dialogue between experimentation and theoretical analysis---engaging the imagination at every turn---and leads through logical frameworks and equations to unexpected insights. The story we recount here begins in the year 1834, when the Scottish Union Canal Society, wary of competition from railways, commissioned a twenty-six-year-old naval engineer  from Edinburgh to look into the possibility of using a steam-powered tug instead of horses to tow barges.

To study how water impedes the propulsion of floating bodies, John Scott Russell undertook fieldwork on the Union Canal that links Edinburgh and Glasgow. Near the village of Hermiston, adjacent to what is now the campus of Heriot-Watt University, he encountered a most remarkable sight, which he described this way:$\,$\cite{ScottRussellX}
\begin{quote}
``I was observing the motion of a boat which was rapidly drawn along a narrow channel by a pair of horses, when the boat suddenly stopped---not so the mass of water in the channel which it had put in motion; it accumulated round the prow of the vessel in a state of violent agitation, then suddenly leaving it behind, rolled forward with great velocity, assuming the form of a large solitary elevation, a rounded, smooth and well-defined heap of water, which continued its course along the channel apparently without change of form or diminution of speed. I followed it on horseback, and overtook it still rolling on at a rate of some eight or nine miles an hour, preserving its original figure some thirty feet long and a foot to a foot and a half in height. Its height gradually diminished, and after a chase of one or two miles I lost it in the windings of the channel. Such, in the month of August 1834, was my first chance interview with that singular and beautiful ph\ae nomenon which I have called the Wave of Translation \ldots'' 
\end{quote}

To learn more about the character of the novel wave, Scott Russell built a long shallow wooden channel---a wave tank---in which he could generate waves of varying amplitudes and measure their properties. An extensive experimental campaign  led him to verify that the wave of translation exhibits an uncanny stability, propagating over considerable distances without changing its form. Ordinary water waves, in contrast, tend to spread out, or to break at the crest. {See Ref.~\citen{carnot_hydrodynamic_soliton} for video of a solitary wave created in a modern wave tank, 18 meters long, at the Institut Carnot de Bourgogne.}  In his comprehensive 1844 report, Scott Russell deduced a simple expression for the propagation velocity,
\begin{equation}
v = \sqrt{g(h + k)}\; , \label{eq:jsr}
\end{equation}
where $g \approx 32\fss \approx 9.8\mss$ is the gravitational acceleration at Earth's surface, $h$ is the depth of the water in repose, and $k$ is the height of the wavecrest over the plane of repose. The solitary waves move swiftly: a lump 0.2\m\ high on a 1-m-deep channel propagates at 12\hbox{ km/h}!

The taller the wave, the faster it moves, so a taller wave will  overtake  a shorter one. As the two waves come together, they disturb each other, but they do not merge, whereas conventional waves would merge or dissipate. {This can be seen fleetingly in the video of a two-collision solitary-wave in the experimental channel of the Institut Carnot, Ref.~\citen{carnot_kdv_collision}.} As we shall see, each disturbance emerges from the interaction with its figure and velocity as before, but with an offset, or phase shift. 

John Scott Russell's formidable forensic work on the wave of translation was not the full extent of his contributions to science and technology. As an obituary notes$\,$\cite{10.1680/imotp.1887.21314}, 
``[h]e was by turns,
schoolmaster, University professor, experimentalist, ship- and
engine-builder, secretary of a scientific society and of the Great
Exhibition of 1851, and finally consulting-engineer and author.''

\section{An Equation of Untold Richness \label{sec:kdv}}
All this is lovely. But is it true? And if it is true, what explains it? Over the next half century, Scott Russell's wave of translation attracted considerable interest---both experimental and analytical---but little consensus. Into this unsettled situation stepped Diederik Johannes Korteweg (1848--1941), Professor of Mathematics at the University of Amsterdam, and his proteg\'e, Gustav de Vries (1866--1934)$\,$\footnote{The Mathematics Institute at the University of Amsterdam is named in their honor. For brief biographical notes, see Ref.~\citen{KdVbios}.}. They open their 1895 treatise, ``On the change of form of long waves
advancing in a rectangular canal, and on a new type of long stationary
waves,'' with these lines$\,$\cite{Korteweg:1895lrm}:
\begin{quote}

``In such excellent treatises on hydrodynamics as those of
Lamb\cite{Lamb} and Basset\cite{Basset}, we find that even when friction is neglected
long waves in a rectangular canal must necessarily
change their form as they advance, becoming steeper in
front and less steep behind. Yet since the investigations
of Boussinesq$\,$\cite{Bouss2}, Lord Rayleigh and [Adh\'{e}mar Jean Claude Barr\'{e} de] Saint-Venant on the solitary
wave, there has been some cause to doubt the truth of
this assertion. Indeed, if the reasons adduced were really
decisive, it is difficult to see why the solitary wave
should make an exception; but even Lord Rayleigh\cite{rayleigh} and
McCowan,\cite{McCowan01071891} who have successfully and thoroughly treated the
theory of this wave, do not directly contradict the statement
in question. They are, as it seems to us, inclined to
the opinion that the solitary wave is only stationary to a
certain approximation.\\

``It is the desire to settle this question definitively
which has led us into the somewhat tedious calculations
which are to be found at the end of our paper. We believe,
indeed, that from them the conclusion may be drawn, that
\emph{in a frictionless liquid there may exist absolutely stationary
waves} and that the form of their surface and the
motion of the liquid below it may be expressed by means of
rapidly convergent series [emphasis added].''
\end{quote}

What emerged from their labors is the nonlinear equation we know as the Korteweg--de Vries (KdV) equation for the disturbance $v(x,t)$, which we write as
\begin{equation}
v_t - 6vv_x + v_{xxx} = 0 . \label{eq:kdv}
\end{equation}
Here $v(x,t)$ is a function of the position $x$ that depends on a time parameter $t$. While the KdV equation first arose in the study of solitary  waves in shallow water channels, it is relevant to a variety of systems that exhibit a balance between nonlinear and dispersive effects---the second and third terms in Eq.~(\ref{eq:kdv}), respectively.

The solution  corresponding to an isolated solitary wave is ($\sech{z} \equiv 1/\cosh{z}$)
\begin{equation}
v(x,t) = 2\kappa_n^2\sech^2[\kappa_n(x - 4\kappa_n^2 t -\xi_n)], \label{eq:onesol}
\end{equation}
where the parameter $\kappa_n^2$ sets the amplitude and speed of the disturbance and $\xi_n$ specifies the location of the disturbance at time $t=0$. The correlation between amplitude and speed of the solitary wave is manifest. In the context of the Schr\"odinger equation, the well given by $-v(x,t)$ corresponds to a reflectionless potential$\,$\footnote{In other words, no reflected wave in the continuum.} supporting a single bound state characterized by binding energy $E_n = - \kappa_n^2$.

\begin{figure}[t]
\centerline{\includegraphics[width=0.3\textheight]{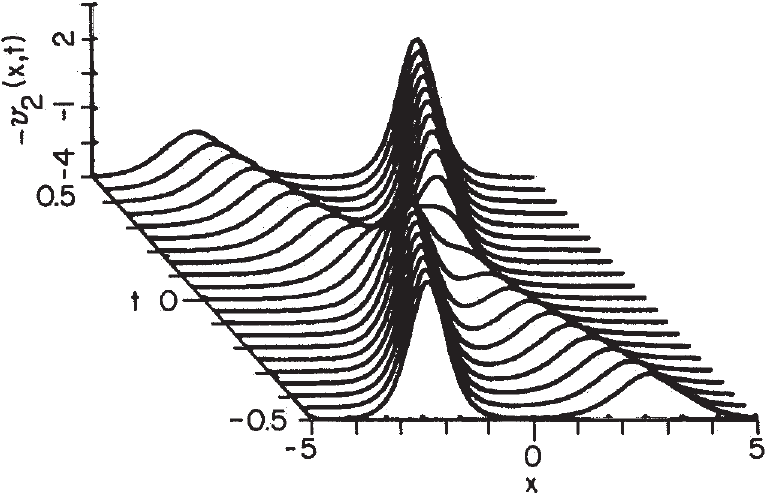}}
\caption{Scattering of two solitary waves, represented by Korteweg--de Vries solitons, from Ref.~\citen{Thacker:1977aq}. \label{fig:2solitons}}
\end{figure}

It is interesting to examine the interaction of two KdV solutions with different amplitudes. As Norman Zabusky and Martin Kruskal observed in their pioneering simulations~\cite{PhysRevLett.15.240}, during the interval when waves overlap, the amplitude of the joint disturbance \emph{decreases}, contrary to what happens when waves overlap linearly. This is illustrated in Figure~\ref{fig:2solitons}. After their encounter, the lumps recover their original figures and continue to propagate at their initial velocities---but with a phase advance. It is as if they were momentarily displaced by the encounter.\footnote{See Figure~1 of Ref.~\citen{Thacker:1977aq} for examples of three-, four-, and five-soliton collisions.}

Zabusky and Kruskal named the stable structures they encountered \emph{solitons.} Their work helped launch a consequential wave of new developments in applied mathematics~\cite{Goda,zbMATH03588211}. The evolution equations that govern solitons exhibit infinitely many conserved quantities, pointing to the existence of many far-from-obvious symmetries that lie behind the remarkable stability of these fascinating lumps~\cite{Palais}.


%

\section{Inverse Scattering: Reflectionless Approximations to Potentials \label{sec:1dform}}
Following the discovery of the $\jpsi$ $(c\bar{c})$ and $\Upsilon$ $(b\bar{b})$ families of heavy mesons in the nineteen-seventies, my colleagues and I were motivated to ask: How and to what extent does the spectrum of a quarkonium system measure the interquark potential? The mathematical problem this question poses had been studied for many years in other contexts, and a rich formalism had grown up about it (cf.\ References 5--11 of Ref.~\citen{Thacker:1977aq}, where full details of our own ``somewhat tedious calculations'' may be found). We found it rewarding to adapt the classic techniques of Gel'fand and Levitan$\,$\cite{Gelfand1955OnTD} and of Kay and Moses$\,$\cite{10.1063/1.1722296}.

The first step is to consider the one-dimensional Schr\"odinger equation,
\begin{equation}
\left(-\frac{\partial^2}{\partial x^2 }+ V(x)\right)\phi(x,k) = k^2\phi(x,k) ,\label{eq:1dschrod}
\end{equation}
where the potential $V(x)$ supports $N$ bound states.
The challenge is to use information about the bound states and scattering in the continuum to determine characteristics of the potential. If the reflection coefficient vanishes in the continuum, then the Gel'fand--Levitan integral equation reduces to a system of $N$ linear  equations for the bound-state wave functions in terms of $2N$ parameters. A $2N$-parameter expression for the potential $V(x)$ also emerges. Half of the needed parameters are determined by the bound-state energies. The remaining $N$ parameters may be fixed either by some explicit piece of information about the bound-state wave functions (i.e., their values at $x=0$), or by imposing the requirement $V(x) = V(-x)$.

When the physical situation permits the continuum to be ignored, the inverse problem is completely and explicitly solved. The result is a symmetric, reflectionless potential that binds $N$ states at arbitrarily adjustable energies. For a strictly confining potential, there is perforce no continuum, but it is interesting to ask how well the inverse scattering formulas work with only partial information, namely the energies of a few low-lying bound states. Implementing this program relies on the now-celebrated connection between the inverse problem for the Schr\"odinger equation and the Korteweg--de Vries equation$\,$\cite{Gardner:1967wc,GGKMCMP,Scott:1973eg}.

To investigate this question, we constructed reflectionless potentials that reproduce the first 1, 2, 3, 4, and 5 levels of familiar power-law potentials and the corresponding wave functions. The components are individual solitons, each supporting a bound state of energy $E_n$, arranged to coicide at time $t=0$. Even having specialized to reflectionless potentials, the reconstructed potentials are not uniquely determined. It is necessary to choose an ionization energy $E_0$ to define the binding energies $\kappa_n^2 \equiv E_0 - E_n$. In specific examples we find that the best $N$-level approximate reconstruction is obtained for $E_0 =(E_N + E_{N+1})/2$. 

\begin{figure}[b]
\centerline{\includegraphics[width=0.3\textheight]{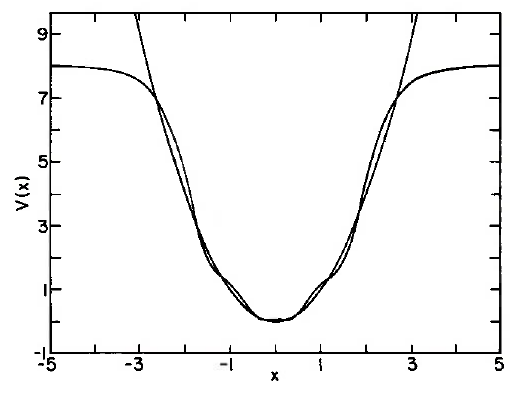}}
\caption{Reflectionless  approximation to the harmonic oscillator potential, $V(x) = x^2$, constructed from four bound states, from Ref.~\citen{Thacker:1977aq}. \label{fig:sho}}
\end{figure}

The 4-level approximation to the harmonic oscillator potential, $V(x) = x^2$, is shown in Figure~\ref{fig:sho}. The fidelity is satisfying, as is the quality of the reconstructed bound-state wave functions. Reflectionless approximations to the linear potential, $V(x) = \lvert x \rvert$, the expected form for the confining potential between a quark and antiquark, are similarly satisfying. 
In both examples, as additional levels are included, the approximation to $V(x)$ is improved locally and is extended to larger values of $\lvert x \rvert$. Compare Figures 3 and 4 of Ref.~\citen{Thacker:1977aq}.

The reconstructed potential is not confining, but approximates $V(x)$ over a range that is roughly delimited by the classical turning point $\lvert x_N \rvert$ of the highest level included, where $V(\pm\lvert x_N \rvert) = E_N$. 
Beyond the classical turning point, $V_N(x)$ asymptotes to the value $E_0$ introduced above. One can see how a many-level sequence of reflectionless potentials would converge to a faithful representation of a confining potential$\,$\cite{Schonfeld:1979cd}.

\section{Probing the Force between Quarks}
We were drawn to the inverse scattering problem to discover how well we might characterize the quark--antiquark interaction in the region of space that governs the spectra of the narrow $\jpsi$ (charm--anticharm, $c\bar{c}$) and $\Upsilon$ (bottom--antibottom, $b\bar{b}$) levels. How compatible would the determinations from the two families be? How would they compare with the potentials motivated by theoretical ideas or inferred from scaling laws for the quarkonium observables?

Within the framework of the nonrelativistic Schr\"odinger equation with a central potential,
\begin{equation}
\left(\frac{-\nabla^2}{2\mu} + V(r)\right)\Psi(\vec{r}) = E\Psi(\vec{r}) , \label{eq:3dschrod}
\end{equation}
where $\mu$ is the reduced mass of the bound system, the procedure for determining $V(r)$ is explicit and essentially unambiguous.~\cite{Thacker:1977ar} 
We confine our attention to the s-wave spin-triplet ($^3\mathrm{S}_1$) quarkonium states below open-flavor threshold. For the one-dimensional case discussed  in \S\ref{sec:1dform}, the reflectionless potential reconstructed from $N$ levels was specified by the $N$ bound-state parameters
\begin{equation}
\kappa_n^2 = 2\mu(E_0 - E_n) .\label{eq:kappadef}
\end{equation}

For s-wave states in three dimensions, the reconstruction is formally similar, with an important difference. In the one-dimensional case, both odd-parity and even-parity states are physically allowed. In three dimensions, the requirement that the radial wave function be finite at the origin $r=0$ means that only the odd-parity solutions correspond to physical s-waves. We must therefore regard \jpsi\ and $\psi^\prime$ as the second and fourth levels of a symmetric one-dimensional potential $V(r) = V(-r)$. The even-parity levels that appear in the one-dimensional problem are interleaved with the physical charmonium states, one below \jpsi, one between \jpsi\ and $\psi^\prime$, and so on.

The values of $\kappa_2, \kappa_4, \ldots$ are determined directly from Eqn.~\ref{eq:kappadef}. The ``unphysical'' parameters $\kappa_1, \kappa_3, \ldots$ do not have the same immediate physical significance, but the wave functions of the physical states depend on them. The square of a $^3\mathrm{S}_1$ wave function at the origin is measured  by the leptonic decay rate as 
\begin{equation}
\lvert\Psi(0)\rvert^2 = \frac{M_{\mathcal{V}}^2}{16\pi\alpha^2 e_Q^2}\Gamma(\mathcal{V}\to e^+e^-), \label{eq:vanrw}
\end{equation}
where $M_{\mathcal{V}}$ is the vector-meson mass and $e_Q$ is the charge of the constituent quark. Quantum corrections are readily included. This piece of information permits the determination of the odd-numbered $\kappa_i$ from experimental data.

Purely for illustration, let us consider reconstructing a potential from information about a single charmonium state, \jpsi. The parameter $\kappa_2$ is given by
\begin{equation}
\kappa_2 = \sqrt{m_c(E_0 - M_{\jpsi})}. \label{eq:kap2}
\end{equation}
From the approximate wave function deduced from the inverse-scattering protocol, we deduce
\begin{equation}
\kappa_1^2 = \kappa_2^2 + \frac{4\pi\lvert\Psi(0)\rvert^2}{\kappa_2}. \label{eq:kap1}
\end{equation}
Explicit, but more complicated, expressions may be derived for the odd-numbered $\kappa_i$ when more physical states are included.

As a first application, we reconstructed potentials from the two narrow charmonium vector states \jpsi\ and $\psi^\prime$. The reflectionless potentials closely resembled the Coulomb + linear and similar potentials then in phenomenological use, and yielded satisfying predictions for the properties of $^3\mathrm{S}_1$ $\Upsilon$ states. As more information accumulated about the properties of $\Upsilon$ states, it became possible to do the exercise in both directions: to predict $\Upsilon$ properties from a potential reconstructed from the charmonium levels, and vice versa\cite{Quigg:1979pi}. The comparison provided constructive evidence that the quark--antiquark potential is flavor-independent.

The possibility of constructing designer (reflectionless) potentials enables us to examine familiar quantum-mechanical results from new perspectives. We see, for example, how band structure requires a periodic potential, and how the injunction against degenerate levels in one-dimensional quantum mechanics materializes\cite{Kwong:1980cu}.


\section{Concluding Remarks\label{sec:conclusion}}
The Korteweg--de Vries equation emerged over many years as a magical result that teaches us far more than anyone might have  anticipated at the outset. Applications and insights for theoretical particle physics are presented in admirable lecture courses by Sidney Coleman$\,$\cite{Coleman1977} and David Tong\cite{Tong2005TASI}. Introductions to the conceptual and technological consequences of solitons in fibers$\,$\cite{fibersolitons,Hasegawa2022Review}, provide a broader view of the importance of the phenomenon first observed by John Scott Russell, now refined and extended by the work of many hands.

The extraordinary scope of the KdV equation evokes the richness of Maxwell's equations, as celebrated by Heinrich Hertz, who demonstrated the existence of electromagnetic radiation:
\begin{quote}
``One cannot study Maxwell's marvelous electromagnetic theory
of light without sometimes having the feeling that these
mathematical formulae have an independent existence and an
intelligence of their own, that they are wiser than we are, wiser
even than their discoverers, that we got more of them than was
originally put into them\cite{001021366}.''
\end{quote}

\section*{Acknowledgments}
I thank my collaborators on studies of inverse-scattering methods applied to quarkonium states. Hank Thacker introduced me to the Korteweg--de Vries equation and its connection to reflectionless potentials. The late Jonathan Rosner was a driving force in all of our investigations. Waikwok Kwong, Peter Moxhay, and Jonathan Schonfeld made unique contributions to pieces of the work. 

Fermilab is operated by Fermi Forward Discovery Group, LLC, under Contract No.\ 89243024CSC000002 with the U.S. Department of Energy, Office of Science, Office of High Energy Physics. Publisher acknowledges the U.S. Government license to provide public access under the DOE Public Access Plan.

%
%

\appendix


\bibliographystyle{ws-ijmpa}
\bibliography{CQKdV}
\end{document}